# Neural network models and deep learning – a primer for biologists

Nikolaus Kriegeskorte[1,2,3,4] and Tal Golan[4]

[1]Department of Psychology, [2]Department of Neuroscience
[3]Department of Electrical Engineering
[4]Zuckerman Mind Brain Behavior Institute, Columbia University
n.kriegeskorte@columbia.edu, tal.golan@columbia.edu

**Originally inspired by neurobiology, deep neural network models have become a powerful tool of machine learning and artificial intelligence. They can approximate functions and dynamics by learning from examples. Here we give a brief introduction to neural network models and deep learning for biologists. We introduce feedforward and recurrent networks and explain the expressive power of this modeling framework and the backpropagation algorithm for setting the parameters. Finally, we consider how deep neural network models might help us understand brain computation.**

## Neural network models of brain function

Brain function can be modeled at many different levels of abstraction. At one extreme, neuroscientists model single neurons and their dynamics in great biological detail. At the other extreme, cognitive scientists model brain information processing with algorithms that make no reference to biological components. In between these extremes lies a model class that has come to be called *artificial neural network* (Rumelhart et al., 1987; LeCun et al., 2015; Schmidhuber, 2015; Goodfellow et al., 2016).

A biological neuron receives multiple signals through the synapses contacting its dendrites and sends a single stream of action potentials out through its axon. The conversion of a complex pattern of inputs into a simple decision (to spike or not to spike) suggested to early theorists that each neuron performs an elementary cognitive function: it reduces complexity by categorizing its input patterns (McCulloch and Pitts, 1943; Rosenblatt, 1958). Inspired by this intuition, artificial neural network models are composed of *units* that combine multiple inputs and produce a single output.

The most common type of unit computes a weighted sum of the inputs and transforms the result nonlinearly. The weighted sum can be interpreted as comparing the pattern of inputs to a reference pattern of weights, with the weights corresponding to the strengths of the incoming connections. The weighted sum is called the *preactivation*. The strength of the preactivation reflects the overall strength of the inputs and, more importantly, the match between the input pattern and the weight pattern. For a given input strength (measured as the sum of squared intensities), the preactivation will be maximal if the input pattern exactly matches the weight pattern (up to a scaling factor).

The preactivation forms the input to the unit's nonlinear activation function. The activation function can be a threshold function (0 for negative, 1 for positive preactivations), indicating whether the match is sufficiently close for the unit to respond (Rosenblatt, 1958). More typically, the activation function is a monotonically increasing function, such as the logistic function (Figure 1) or a rectifying nonlinearity, which outputs the preactivation if it is positive and zero otherwise. These latter activation functions have non-zero derivatives (at least over the positive range of preactivations). As we will see below, non-zero derivatives make it easier to optimize the weights of a network.

The weights can be positive or negative. Inhibition, thus, need not be relayed through a separate set of inhibitory units, and neural network models typically do not respect Dale's law (which states that a neuron performs the same chemical action at all of its synaptic connections to other neurons, regardless of the identity of the target cell). In addition to the weights of the incoming connections, each unit has a bias parameter: the bias is added to the preactivation, enabling the unit to shift its nonlinear activation function horizontally, for example moving the threshold to the left or right. The bias can be understood as a weight for an imaginary additional input that is constantly 1.

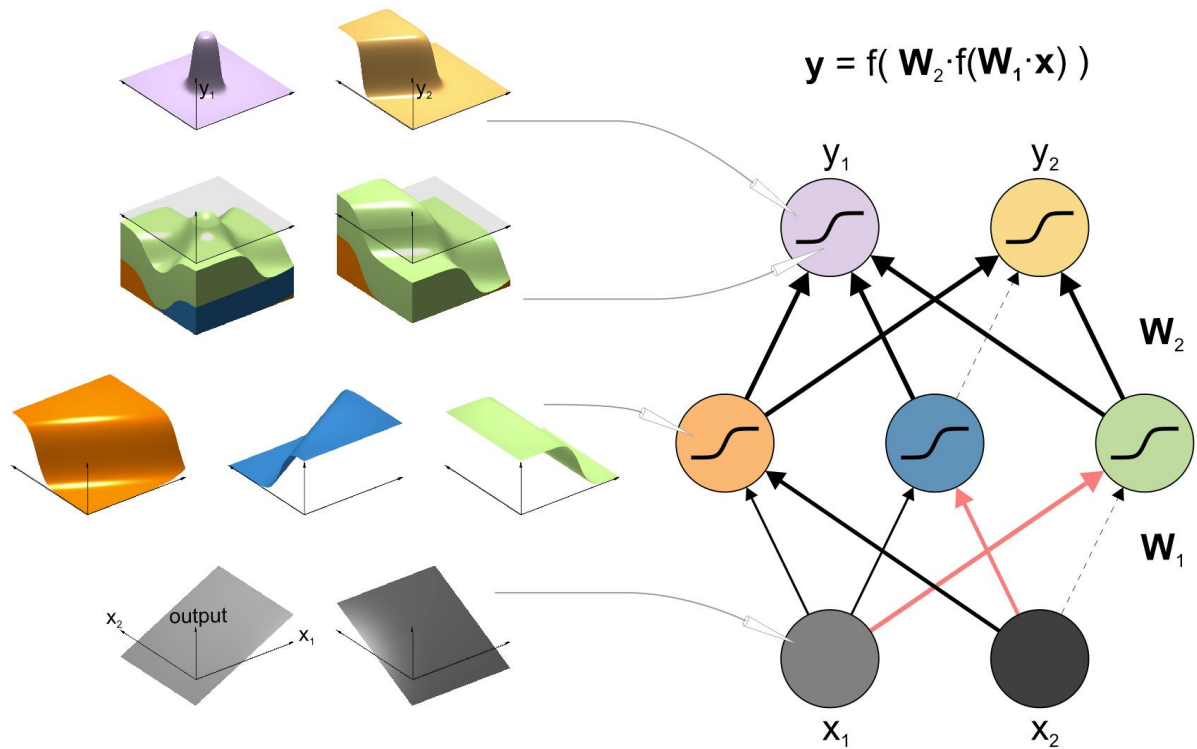

$$y = f(\ W_2 \cdot f(W_1 \cdot x)\ )$$

**Figure 1 | Function approximation by a feedforward neural network.** A feedforward neural network with two input units (bottom), three hidden units (middle), and two output units (top). The input patterns form a two-dimensional space. The hidden and output units here use a sigmoid (logistic) activation function. Surface plots on the left show the activation of each unit as a function of the input pattern (horizontal plane spanned by inputs $x_1$ and $x_2$). For the output units, the preactivations are shown below the output activations. For each unit, the weights (arrow thickness) and signs (black, positive; red, negative) of the incoming connections control the orientation and slope of the activation function. The output units combine the nonlinear ramps computed by the hidden units. Given enough hidden units, a network of this type can approximate any continuous function to arbitrary precision.

## Neural networks are universal approximators

Units can be assembled into networks in many different configurations. A single unit can serve as a linear discriminant of its input patterns. A set of units connected to the same set of inputs can detect multiple classes, with each unit implementing a different linear discriminant. For a network to discriminate classes that are not linearly separable in the input signals, we need an intermediate layer between input and output units, called a *hidden layer* (Figure 1).

If the units were linear — outputting the weighted sum directly, without passing it through a nonlinear activation function — then the output units reading out the hidden units would compute weighted sums of weighted sums and would, thus, themselves be limited to weighted sums of the inputs. With nonlinear activation functions, a hidden layer makes the network more expressive, enabling it to approximate nonlinear functions of the input, as illustrated in Figure 1.

A feedforward network with a single hidden layer (Figure 1) is a flexible approximator of functions that link the inputs to the desired outputs. Typically, each hidden unit computes a nonlinear ramp, for example sigmoid or rectified linear, over the input space. The ramp rises in the direction in input space that is defined by the vector of incoming weights. By adjusting the weights, we can rotate the ramp in the desired direction. By scaling the weights vector, we can squeeze or stretch the ramp to make it rise more or less steeply. By adjusting the bias, we can shift the ramp forward or backward. Each hidden unit can be independently adjusted in this way.

One level up, in the output layer, we can linearly combine the outputs of the hidden units. As shown in Figure 1, a weighted sum of several nonlinear ramps produces a qualitatively different continuous function over the input space. This is how a hidden layer of linear–nonlinear units enables the approximation of functions very different in shape from the nonlinear activation function that provides the building blocks.

It turns out that we can approximate any continuous function to any desired level of precision by allowing a sufficient number of units in a single hidden layer (Cybenko, 1989; Hornik et al., 1989). To gain an intuition of why this is possible, consider the left output unit ($y_1$) of the network in Figure 1. By combining ramps overlapping in a single region of the input space, this unit effectively selects a single compact patch. We could tile the entire input space with sets of hidden units that select different patches in this way. In the output layer, we could then map each patch to any desired output value. As we move from one input region to another, the network would smoothly transition between the different output values. The precision of such an approximation can always be increased by using more hidden units to tile the input space more finely.

## Deep networks can efficiently capture complex functions

A feedforward neural network is called '*deep*' when it has more than one hidden layer. The term is also used in a graded sense, in which the depth denotes the number of layers. We have seen above that even *shallow* neural networks, with a single hidden layer, are universal function approximators. What, then, is the advantage of *deep* neural networks?

Deep neural networks can re-use the features computed in a given hidden layer in higher hidden layers. This enables a deep neural network to exploit compositional structure in a function, and to approximate many natural functions with fewer weights and units (Lin et al., 2017; Rolnick and Tegmark, 2017). Whereas a shallow neural network must piece together the function it approximates, like a lookup table (although the pieces overlap and sum), a deep neural network can benefit from its hierarchical structure. A deeper architecture can increase the precision with which a function can be approximated on a fixed budget of parameters and can improve the generalization after learning to new examples.

Deep learning refers to the automatic determination of parameters deep in a network on the basis of experience (data). Neural networks with multiple hidden layers are an old idea and were a popular topic in engineering and cognitive science in the 1980s (Rumelhart et al., 1987). Although the advantages of deep architectures were understood in theory, the method did not realize its potential in practice, mainly because of insufficient computing power and data for learning. Shallow machine learning techniques, such as support vector machines (Cortes and Vapnik, 1995; Schölkopf and Smola, 2002), worked better in practice and also lent themselves to more rigorous mathematical analysis. The recent success of deep learning has been driven by a rise in computing power — in particular the advent of graphics processing units, GPUs, specialized hardware for fast matrix–matrix multiplication — and web-scale data sets to learn from. In addition, improved techniques for pretraining, initialization, regularization, and normalization, along with the introduction of rectified linear units, have all helped to boost performance. Recent work has explored a wide variety of feedforward and recurrent network architectures, improving the state-of-the-art in several domains of artificial intelligence and establishing deep learning as a central strand of machine learning in the last few years.

The function that deep neural networks are trained to approximate is often a mapping from input patterns to output patterns, for example classifying natural images according to categories, translating sentences from English to French, or predicting tomorrow's weather from today's measurements. When the cost minimized by training is a measure of the mismatch between the network's outputs and desired outputs (that is, the 'error'), for a training set of example cases, the training is called supervised. When the cost minimized by training does not involve prespecified desired outputs for a set of example inputs, the training is called unsupervised.

Two examples of unsupervised learning are autoencoders and generative adversarial networks. Autoencoder networks (Rumelhart et al., 1986; Hinton and Salakhutdinov, 2006) learn to transform input patterns into a compressed latent representation by exploiting inherent statistical structure. Generative adversarial networks (Goodfellow et al., 2014) operate in the opposite direction, transforming random patterns in a latent representation into novel, synthetic examples of a category, such as fake images of bedrooms. The generator network is trained concurrently with a discriminator network that learns to pick out the generator's fakes among natural examples of the category. The two adversarial networks boost each other's performance by posing increasingly difficult challenges of counterfeiting and detection to each other. Deep neural networks can also be trained by reinforcement (deep reinforcement learning), which has led to impressive performance at playing games and robotic control (Mnih et al., 2015).

# Deep learning by backpropagation

Say we want to train a deep neural network model with supervision. How can the connection weights deep in the network be automatically learned? The weights are randomly initialized and then adjusted in many small steps to bring the network closer to the desired behavior. A simple approach would be to consider random perturbations of the weights and to apply them when they improve the behavior. This evolutionary approach is intuitive and has recently shown promise (Such et al., 2017; Stanley et al., 2019), but it is not usually the most efficient solution. There may be millions of weights, spanning a search space of equal dimension. It takes too long in practice to find directions to move in such a space that improve performance. We could wiggle each weight separately, and determine if behavior improves. Although this would enable us to make progress, adjusting each weight would require running the entire network many times to assess its behavior. Again, progress with this approach is too slow for many practical applications.

In order to enable more efficient learning, neural network models are composed of *differentiable* operations. How a small change to a particular weight affects performance can then be computed as the partial derivative of the error with respect to the weight. For different weights in the same model, the algebraic expressions corresponding to their partial derivatives share many terms, enabling us to efficiently compute the partial derivatives for all weights.

For each input, we first propagate the activation forward through the network, computing the activation states of all the units, including the outputs. We then compare the network's outputs with the desired outputs and compute the cost function to be minimized (for example, the sum of squared errors across output units). For each unit, we then compute how much the cost would drop if the activation changed slightly. This is the *sensitivity* of the cost to a change of activation of each output unit. Mathematically, it is the partial *derivative* of the cost with respect to the each activation. We then proceed backwards through the network propagating the cost derivatives (sensitivities) from the activations to the preactivations and through the weights to the activations of the layer below. The sensitivity of the cost to each of these variables depends on the sensitivities of the cost to the variables downstream in the network. *Backpropagating* the derivatives through the network by applying the chain rule provides an efficient algorithm for computing all the partial derivatives (Werbos, 1982).

The critical step is computing the partial derivative of the cost with respect to each weight. Consider the weight of a particular connection (red arrow in Figure 2). The connection links a source unit in one layer to a target unit in the next layer. The influence of the weight on the cost for a given input pattern depends on how active the source unit is. If the source unit is off for the present input pattern, then the connection has no signal to transmit and its weight is irrelevant to the output the network produces for the current input. The activation of the source unit is *multiplied* with the weight to determine its contribution to the preactivation of the target unit, so the source activation is one factor determining the influence of the weight on the cost. The other factor is the sensitivity of the cost to the preactivation of the target unit. If the preactivation of the target unit had no influence on the cost, then the weight would have no influence, either. The derivative of the cost with respect to the weight is the product of its source unit's activation and its target unit's influence on the cost.

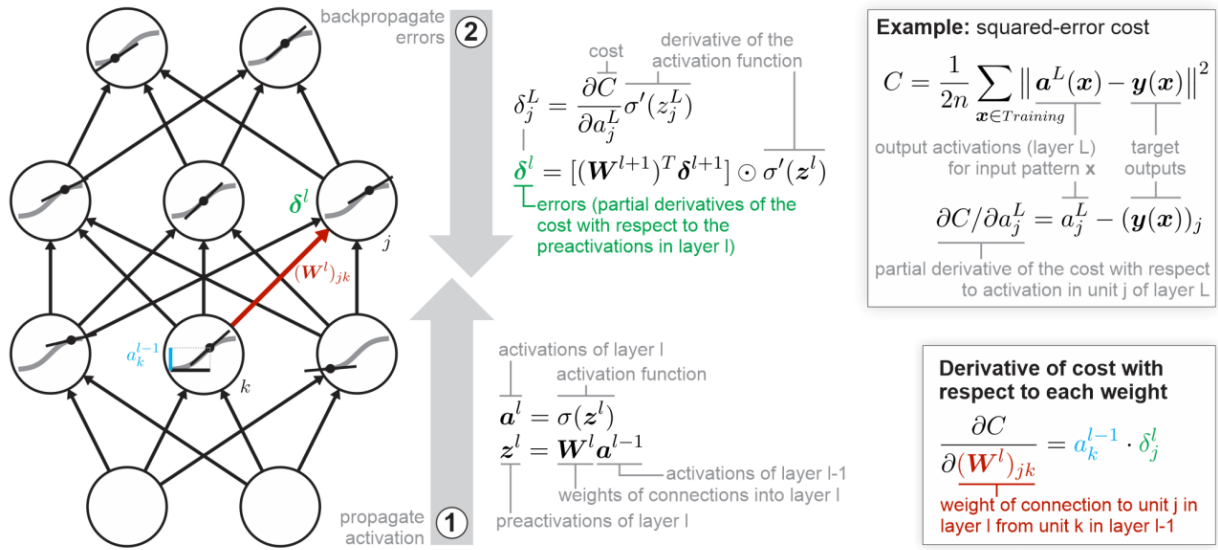

**Figure 2 | The backpropagation algorithm.** Backpropagation is an efficient algorithm for computing how small adjustments to the connection weights affect the cost function that the network is meant to minimize. A feedforward network with two hidden layers is shown as an example. First, the activations are propagated in the feedforward direction (upward). The activation function (gray sigmoid) is shown in each unit (circle). In the context of a particular input pattern (not shown), the network is in a particular activation state, indicated by the black dots in the units (horizontal axis: preactivation; vertical axis: activation). Second, the derivatives of the cost function (squared-error cost shown on the right) are propagated in reverse (downward). In the context of the present input pattern, the network can be approximated as a linear network (black lines indicating the slope of the activation function). The chain rule defines how the cost (the overall error) is affected by small changes to the activations, preactivations, and weights. The goal is to compute the partial derivative of the cost with respect to each weight (bottom right). Each weight is then adjusted in proportion to how much its adjustment reduces the cost. The notation roughly follows Nielsen (2015), but we use bold symbols for vectors and matrices. The symbol ⊙ denotes element-wise multiplication (Hadamard product).

We adjust each weight in the direction that reduces the cost (the error) and by an amount proportional to the derivative of the cost with respect to the weight. This process is called *gradient descent*, because it amounts to moving in the direction in weight space in which the cost declines most steeply. To help our intuition, let us consider two approaches we might take. First, consider the approach of taking a step to reduce the cost *for each individual training example*. Gradient descent will make a *minimal and selective adjustments* to reduce the error, which makes sense as we do not want learning from the current example to interfere with what we've learned from other examples. However, our goal is to reduce the *overall error*, which is defined as the sum of the errors across all examples. So second, consider the approach of summing up the error surfaces (or, equivalently, the gradients) across all examples before taking a step. We can still only take a small step, because the error surface is nonlinear and so the gradient will change as we move away from the point about which we linearized the network.

In practice, the best solution is to use small *batches of training examples* to estimate the gradient before taking a step. Compared to the single-example approach, this gives us a more stable sense of direction. Compared to the full-training-set approach, it greatly reduces the computations required to take a step. Although the full-training-set approach gives exact gradients for the training-set error, it still does not enable us to take large steps, because of the nonlinearity of the error function. Using batches is a good compromise between stability of the gradient estimate and computational cost. Because the gradient estimate depends on the random sample of examples in the current batch, the method is called *stochastic* gradient descent (SGD). Beyond the motivation just given, the stochasticity is thought to contribute also to finding solutions that generalize well beyond the training set (Poggio and Liao, 2017).

The cost is not a convex function of the weights, so we might be concerned about getting stuck in local minima. However, the high dimensionality of weight space turns out to be a blessing (not a curse) for gradient descent: there are many directions to escape in, making it unlikely that we will ever find ourselves trapped, with the error surface rising in all directions (Kawaguchi, 2016). In practice, it is saddle points (where the gradient vanishes) that pose a greater challenge than local minima (Dauphin et al., 2014). Moreover, the cost function typically has many symmetries, with any given set of weights having many computationally equivalent twins (that is, the model computes the same overall function for different parameter settings). As a result, although our solution may be one local minimum among many, it may not be a poor local minimum: It may be one of many similarly good solutions.

## Recurrent neural networks are universal approximators of dynamical systems

So far we have considered feedforward networks, whose directed connections do not form cycles. Units can also be configured in recurrent neural networks (RNNs), where activity is propagated in cycles, as is the case in brains (Dayan and Abbott, 2001; Goodfellow et al., 2016). This enables a network to recycle its limited computational resources over time and perform a deeper sequence of nonlinear transformations. As a result, RNNs can perform more complex computations than would be possible with a single feedforward sweep through the same number of units and connections.

For a given state space, a suitable RNN can map each state to any desired successor state. RNNs, therefore, are universal approximators of dynamical systems (Schäfer and Zimmermann, 2006). They provide a universal language for modeling dynamics, and one whose components could plausibly be implemented with biological neurons.

Much like feedforward neural networks, RNNs can be trained by backpropagation. However, backpropagation must proceed through the cycles in reverse. This process is called backpropagation through time. An intuitive way to understand an RNN and backpropagation through time is to 'unfold' the RNN into an equivalent feedforward network (Figure 3). Each layer of the feedforward network represents a timestep of the RNN. The units and weights of the RNN are replicated for each layer of the feedforward network. The feedforward network, thus, shares the same set of weights across its layers (the weights of the recurrent network).

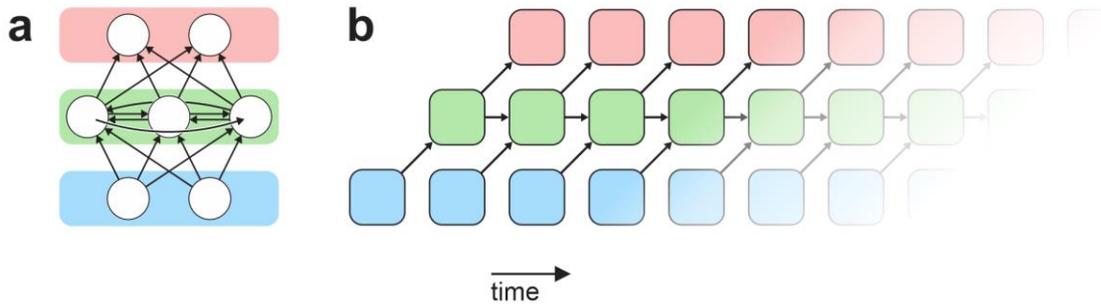

**Figure 3 | Recurrent neural networks.** (a) A recurrent neural network models with two input units (in blue box), three hidden units (green box), and two output units (pink box). The hidden units here are fully recurrently connected: each sends its output to both other units. The arrows represent scalar weights between particular units. (b) Equivalent feedforward network. Any recurrent neural network can be unfolded along time as a feedforward network. To this end, the units of the recurrent neural network (blue, green, pink sets) are replicated for each time step. The arrows here represent weights matrices between sets of units in the colored boxes. For the equivalence to hold, the feedforward network has to have a depth matching the number of time steps that the recurrent network is meant to run for. Unfolding leads to a representation that is less concise, but easier to understand and often useful in software implementations of recurrent neural networks. Training of the recurrent model by backpropagation through time is equivalent to training of the unfolded model by backpropagation.

For tasks that operate on independent observations (for example, classifying still images), the recycling of weights can enable an RNN to perform better than a feedforward network with the same number of parameters (Spoerer et al., 2017). However, RNNs really shine in tasks that operate on streams of dependent observations. Because RNNs can maintain an internal state (memory) over time and produce dynamics, they lend themselves to tasks that require temporal patterns to be recognized or generated. These include the perception speech and video, cognitive tasks that require maintaining representations of hidden states of the agent (such as goals) or the environment (such as currently hidden objects), linguistic tasks like the translation of text from one language into another, and control tasks at the level of planning and selecting actions, as well as at the level of motor control during execution of an action under feedback from the senses.

## Deep neural networks provide abstract process models of biological neural networks

Cognitive models capture aspects of brain information processing, but do not speak to its biological implementation. Detailed biological models can capture the dynamics of action potentials and the spatiotemporal dynamics of signal propagation in dendrites and axons. However, they have only had limited success in explaining how these processes contribute to cognition. Deep neural network models, as discussed here, strike a balance, explaining feats of perception, cognition, and motor control in terms of networks of units that are highly abstracted, but could plausibly be implemented with biological neurons.

For engineers, artificial deep neural networks are a powerful tool of machine learning. For neuroscientists, these models offer a way of specifying mechanistic hypotheses on how cognitive functions may be carried out by brains (Kriegeskorte and Douglas, 2018; Kietzmann et al., 2019; Storrs and Kriegeskorte, in press). Deep neural networks provide a powerful language for expressing information-processing functions. In certain domains, they already meet or surpass human-level performance (for example, visual object recognition and board games) while relying exclusively on operations that are biologically plausible.

Neural network models in engineering have taken inspiration from brains, far beyond the general notion that computations involve a network of units, each of which nonlinearly combines multiple inputs to compute a single output (Kriegeskorte, 2015; Yamins and DiCarlo, 2016). For example, convolutional neural networks (Fukushima and Miyake, 1982; LeCun et al., 1989), the dominant technology in computer vision, use a deep hierarchy of retinotopic layers whose units have restricted receptive fields. The networks are convolutional in that weight templates are automatically shared across image locations (rendering the computation of a feature map's preactivations equivalent to a convolution of the input with the weight template). Although the convolutional aspect may not capture an innate characteristic of the primate visual system, it does represent an idealization of the final product of development and learning in primates, where qualitatively similar features are extracted all over retinotopic maps at early stages of processing. Across layers, these networks transform a visuospatial representation of the image into a semantic representation of its contents, successively reducing the spatial detail of the maps and increasing the number of semantic dimensions (Figure 4).

The fact that a neural network model was inspired by some abstract features of biology and that it matches overall human or animal performance at a task, does not make it a good model of how the human or animal brain performs the task. However, we can compare neural network models to brains in terms of detailed patterns of behavior, such as errors and reaction times for particular stimuli. Moreover, we can compare the internal representations in neural networks to those in brains.

In the 'white-box' approach, we evaluate a model by looking at its internal representations. Neural network models form the basis for predicting representations in different brain regions for a particular set of stimuli (Diedrichsen and Kriegeskorte, 2017). One approach is called *encoding models*. In encoding models, the brain activity pattern in some functional region is predicted using a linear transformation of the representation in some layer of the model (Kay et al., 2008; Mitchell et al., 2008). In another approach, called *representational similarity analysis* (Kriegeskorte et al., 2008; Nili et al., 2014; Kriegeskorte and Diedrichsen, 2016), each representation in brain and model is characterized by a representational dissimilarity matrix. Models are evaluated according to their ability to explain the representational dissimilarities across pairs of stimuli. A third approach is *pattern component modeling* (Diedrichsen et al., 2011, 2017), where representations are characterized by the second moment of the activity profiles.

Recent results from the domain of visual object recognition indicate that deep convolutional neural networks are the best available model of how the primate brain achieves rapid recognition at a glance, although they do not explain all of the explainable variance in neuronal responses. (Cadieu et al., 2014; Khaligh-Razavi and Kriegeskorte, 2014; Yamins et al., 2014; Güçlü and van Gerven, 2015; Cichy et al., 2016; Eickenberg et al., 2017; Wen et al., 2017; Nayebi et al., 2018), and yet multiple functional incompatibilities have already been reported (Szegedy et al., 2013; Geirhos et al., 2017; Jo and Bengio, 2017; Rajalingham et al., 2018).

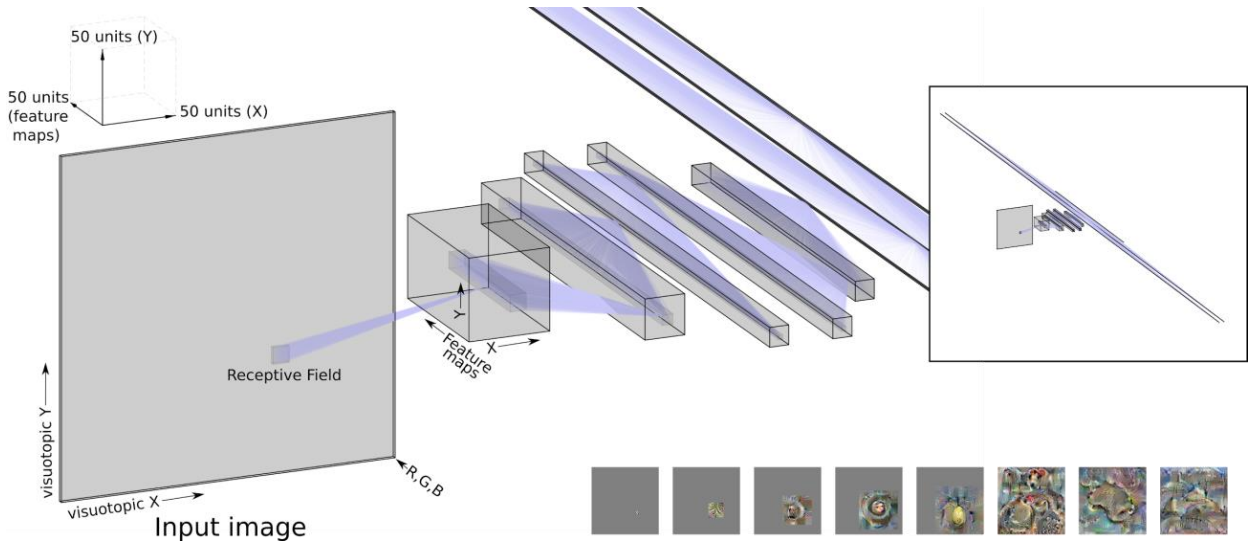

**Figure 4 | Deep convolutional feedforward neural networks.** The general structure of Alexnet, a convolutional deep neural network architecture which had a critical role in bringing deep neural networks into the spotlight. Unlike the visualization in the original report on this model, here the tensors' dimensions are drawn to scale, so it is easier to appreciate how the convolutional deep neural network gradually transforms the input image from a spatial to semantic representation. For sake of simplicity, we did not visualize the pooling operations, as well as the splitting of some of these layers between two GPUs. The leftmost box is the input image, (a tensor of the dimensions 227×227×3, where 227 is the length of the square input-image edges and three is the number of color components). It is transformed by convolution into the first layer (second box from the left), a tensor with smaller spatial dimensions (55×55) but a larger number of feature maps (96). Each feature map in this tensor is produced by a convolution of the original image with a particular 11×11×3 filter. Therefore, the preactivation of each unit in this layer is a linear combination of one rectangular receptive field in the image. The boundaries of such a receptive field are visualized as a small box within the image tensor. In the next, second layer, the representation is even more spatially smaller (27×27) but richer with respect of the number of feature maps (256). Note that from here and onwards, each feature is not a linear combination of pixels but a linear combination of the previous layer's features. The sixth layer (see the small overview inset at the top-right) combines all feature maps and locations of the fifth layer to yield 4096 different scalar units, each with its own unrestricted input weights vector. The final eighth layer has 1000 units, one for each output class. The eight images on the bottom were produced by gradually modifying random noise images so excite particular units in each of the eight layers (Erhan et al., 2009). The rightmost image was optimized to activate the output neuron related to the class 'Mosque'. Importantly, these are only local solutions to the activation-maximization problem. Alternative activation-maximizing images may be produced by using different starting conditions or optimization heuristics.

In the 'black-box' approach, we evaluate a model on the basis of its behavior. We can reject models for failing to explain detailed patterns of behavior. This has already helped reveal some limitations of convolutional neural networks, which appear to behave differently from humans under noisy conditions (Geirhos et al., 2017) and to show different patterns of failures across exemplars (Rajalingham et al., 2018).

Deep neural networks bridge the gap between neurobiology and cognitive function, providing an exciting framework for modeling brain information processing. Theories of how the brain computes can now be subjected to rigorous tests by simulation. Our theories, and the models that implement them, will evolve as we learn to explain the rich measurements of brain activity and behavior provided by modern technologies in animals and humans.